\documentclass[letterpaper, 10 pt, conference]{ieeeconf}  

\IEEEoverridecommandlockouts                              
\usepackage{graphics} 
\usepackage{epsfig} 
\usepackage{mathptmx} 
\usepackage{times} 

\usepackage{svg}
\usepackage{amsmath} 
\usepackage{amssymb}  
\usepackage{cite}
\usepackage{graphicx}
\usepackage{booktabs}
\usepackage{graphicx}
\usepackage{amsmath}

\title{\LARGE \bf
Short Term Blood Glucose Prediction based on \\ Continuous Glucose Monitoring Data}

\author{Ali Mohebbi$^{1,2,3}$, Alexander R. Johansen$^{1}$, Nicklas Hansen$^{1}$,  Peter E. Christensen$^{1}$ \\ Jens M. Tarp$^{4}$, Morten L. Jensen$^{3}$, Henrik Bengtsson$^{2}$, Morten M{\o}rup$^{1}$ 
\thanks{$^{1}$Department of Applied Mathematics and Computer Science, Cognitive Systems, Technical University of Denmark (DTU), Denmark, {\tt\small xaim@novonordisk.com /} {\tt\small mmor@dtu.dk}}
\thanks{$^{2}$ Novo Nordisk A/S, Device R\&D, Hillerød, Denmark}
\thanks{$^{3}$ Novo Nordisk A/S, Medical \& Science, Søborg, Denmark}
\thanks{$^{4}$ Novo Nordisk A/S, Data Analytics, Søborg, Denmark}}

\begin{document}

\maketitle
\thispagestyle{empty}
\pagestyle{empty}

\begin{abstract}
Continuous Glucose Monitoring (CGM) has enabled important opportunities for diabetes management. This study explores the use of CGM data as input for digital decision support tools. We investigate how Recurrent Neural Networks (RNNs) can be used for Short Term Blood Glucose (STBG) prediction and compare the RNNs to conventional time-series forecasting using Autoregressive Integrated Moving Average (ARIMA). A prediction horizon up to 90 min into the future is considered. In this context, we evaluate both population-based and patient-specific RNNs and contrast them to patient-specific ARIMA models and a simple baseline predicting future observations as the last observed. We find that the population-based RNN model is the best performing model across the considered prediction horizons without the need of patient-specific data. This demonstrates the potential of RNNs for STBG prediction in diabetes patients towards detecting/mitigating severe events in the STBG, in particular hypoglycemic events. However, further studies are needed in regards to the robustness and practical use of the investigated STBG prediction models. 
\end{abstract}
\section{INTRODUCTION}
People with insulin requiring diabetes are challenged to achieve and maintain metabolic control. One key challenge is that insulin is a drug with a very narrow therapeutic index, i.e. a short range between having too high blood glucose (BG) levels (hyperglycemia)  and dangerously low BG levels (hypoglycemia). In this context, identifying the optimal dose size and timing of insulin doses are critical. For patients who make meal time injections, adjusting the individual meal dose to the meal at hand, potentially making a correction bolus or eating correction carbohydrates, is part of managing their diabetes.\\ 
\indent
Digital tools to support the management of diabetes are becoming available \cite{tech}, and often contain some predictive capabilities. For instance, \textit{Short Term} BG (STBG) predictions can benefit patients on insulin treatment who are at risk of hypoglycemia. Due to the narrow therapeutic index of insulin and the risk of prompting an adverse insulin treatment behaviour, the predictive capabilities of current and future digital tools must be as precise as possible given the available data. Obviously, predictions of future BG levels depend on the amount and quality of the past BG level data. Furthermore, the need for insulin is dynamically impacted by several factors such as meal size/composition and physical activity level. Hence, predictions of future BG levels benefit from having such data available as well \cite{cobelli}. However, although access to more data sources (carbohydrates, exercise, dose size) add further information and can improve the prediction performance, the availability of such data suffer from it typically having to be actively collected and therefore are either completely lacking or in low quality. As a result, BG predictions have to perform well even when only BG data are available. Another issue is that there may be situations where the immediate historic data are not available, e.g. when a patient is using the tool for the first time or if data has not been generated or collected during a weekend or vacation. Therefore, BG predictions also have to perform well or fail safe in situations where the immediate historic data is limited.  \\
\indent
The use of Continuous Glucose Monitoring (CGM) is increasing and unlocks new possible approaches to diabetes therapy \cite{consensus, mohebbi, mohebbi2}. With CGM, an abundance of real-time BG measurements become available for patients with type 1 diabetes (T1D) and progressed type 2 diabetes (T2D). The question is how the CGM data can be optimally leveraged to prepare future decision support tools. For example, access to large quantities of CGM data creates an opportunity to learn structural properties of the CGM signal, which may be used for predicting future BG values. Thus, optimised predictive modeling could be leveraged to improve the opportunities for patients to make therapeutic decisions based on forecasted BG levels. \\
\indent
Several well-established predictive methods exist in time-series analysis, e.g. the Autoregressive Integrated Moving Average (ARIMA), which is a robust linear approach commonly used in forecasting tasks \cite{ARIMA, stbg_ref}. A recent addition to predictive methods for time-series is the use of Recurrent Neural Networks (RNNs) in Deep Learning (DL). RNNs have proven effective for a variety of sequence modeling and forecasting tasks, most notably in language and speech \cite{AI}. RNNs are typically implemented with Long Short Term Memory (LSTM) units that provides a gating mechanism enabling the RNNs to preserve in memory events further in the past \cite{lstm, graves}. In this context, previous studies have indicated the benefits of RNNs for STBG prediction \cite{rcnn, stbg_ref}. Additionally, motivated by recent results in Transfer Learning (TL), we would like to investigate the ability of transferring models learned from one task to another \cite{transfer}. The question is whether there would be benefits of deploying the more advanced predictive methods for BG predictions compared to traditional methods such as ARIMA. \\
\indent
The specific research questions in the study are: 
\begin{enumerate}
    \item \emph{Does a population based LSTM model generalize well to unseen patients, i.e. if CGM trends acquired from a population can enable STBG prediction on individual patients?
    \item Does increasingly added patient-specific CGM data to a pretrained population based LSTM (i.e., using TL) improve predictive performance on individual patients?
    \item How does the LSTM based solely on patient-specific data perform when trained on limited CGM data?}
\end{enumerate}
\indent
In order to answer the specific research questions, this study systematically investigates LSTM based RNNs for STBG prediction up to 90 min horizon into the future in comparison to ARIMA. Additionally, the study explores the feasibility/application of TL in this context. The models outlined below are compared in this study: %
\begin{enumerate}
    \item Population based LSTM (pretraining),
    \item Finetuned Patient-specific LSTM (with pretraining),
    \item Patient-specific LSTM (without pretraining),
    \item Patient-specific ARIMA.
\end{enumerate}

\section{METHODS}
\begin{figure*}[t!]
    \centering
    \includegraphics[scale=0.71]{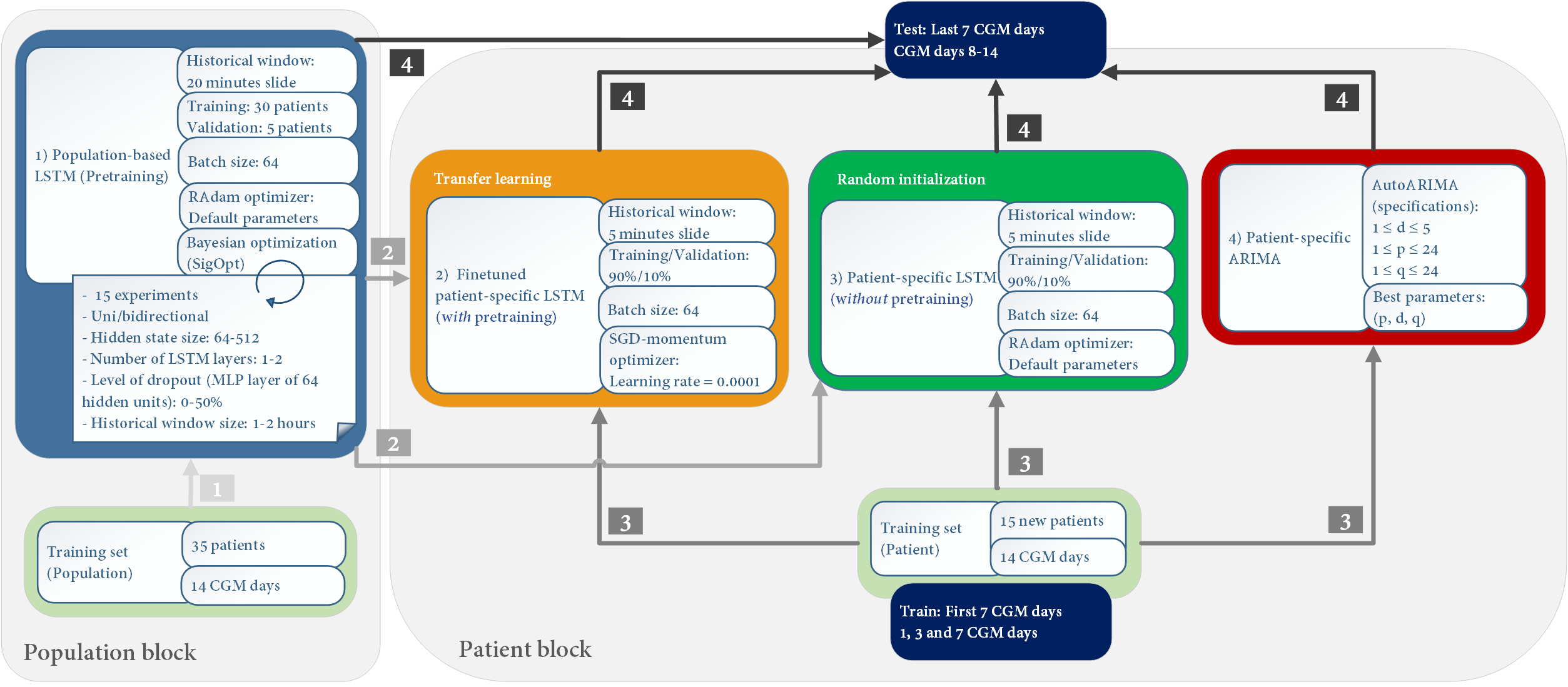}
    \caption{Flowdiagram of the experimental setup including both the population-based and patient-specific models presented by two different blocks. Furthermore, the flow of information within/between blocks and in which order are depicted by the numbered squares. All model specifications for both LSTMs and ARIMA are outlined with corresponding color-coded boxes within the two blocks. Blue box: Population-based LSTM (pretraining), Orange box: Finetuned patient-specific LSTM (with pretraining), Green box: Patient-specific LSTM (without pretraining), Red box: Patient-specific ARIMA.}
    \label{fig:flowdiagram}
\end{figure*}

In this section, a comprehensive description of the included CGM data, examined models, and experimental setup is provided. An overview is supplied in Fig. \ref{fig:flowdiagram}.
\subsection{CGM Data}
The CGM data used in this study are acquired from the
Cornerstones4Care\textsuperscript{\textregistered} platform powered by Glooko which is a diabetes management application capturing real-world CGM data. 14 consecutive days of CGM data (henceforth referred to as a \textit{period}) are acquired from 50 patients with diabetes with readings every 5 minutes (i.e. 288 BG readings per day). However, CGM data from real-life patients often have missing values (or \textit{gaps}) caused either by sensor or human error. Inspired by the international CGM consensus, only CGM days and periods with more than 70\% available BG readings are included \cite{consensus}. However, an average of 96\% available BG readings is present for the included periods. Furthermore, no information exist about the diabetes type, demographics or treatment regimens for the included patients.
\subsection{ARIMA}
ARIMA models are considered some of the most flexible and popular autoregressive techniques for continuous time-series forecasting \cite{ARIMA}. ARIMA is specified by 3 parameters: $p$, $d$, and $q$. Here, $p$ refers to the number of autoregressive terms, $d$ represents the number of nonseasonal differences, and $q$ depicts the number of lagged forecast errors in the prediction equation. When working with a data sequence $x_1, x_2, \dots x_n$, we can describe ARIMA as
\begin{equation}
x'_t = c + \phi_1 x'_{t-1} + \dots + \phi_p x'_{t-p} + \\ \epsilon_t + \theta_1 \epsilon_{t-1} + \theta_q \epsilon_{t-q},
\label{eq:arima}
\end{equation}
where \textbf{$x'_t$} denotes the differenced time-series, which has been differenced $d$ times, $c$ is a constant, \textbf{$\phi$} is the AR($p$) component, \textbf{$\theta$} is the MA($q$) component, and $\epsilon$ is the lagged forecast errors.
\subsection{Long Short-Term Memory Network (LSTM)}
RNN is a type of DL architecture which repeatedly uses the same weights along a sequence of data. The LSTM cell extends the recurrent computation of an RNN with a gating mechanism. This gating mechanism allows the LSTM to represent relationships between data points far apart \cite{lstm, graves}. \\
\indent
The CGM data is modeled as a one-dimensional time series, i.e. the BG value at each time step, see Eq. \ref{eq:x}.
To limit the complexity, we only allowed the LSTM to observe the $k$ preceding BG values (denoted as the historical window) before predicting the next BG value.
\begin{align}
    \hat{x}_{t} = f(x_{t-1}, x_{t-2} \dots, x_{t-k}),
    \label{eq:x}
\end{align}
where $x_i$ is a single BG measurement, $\hat{x}_i$ is the predicted BG value, $k$ corresponds to the historical window size, and $f$ is an LSTM based function with a multilayered perceptron (MLP) resulting in one continuous value, $\hat{x}_t$.
The network architecture, $f$, is defined as follows,
\begin{align}
    \hat{x}_{t} = f(x_{t-1}, x_{t-2} \dots, x_{t-k}) &= \textbf{w}^\top\textbf{g}_t\\
    \textbf{g}_t &= a(\textbf{U}\textbf{h}_t+\textbf{b})\\
    \textbf{h}_t &= \text{LSTM}(x_{t-1}, \textbf{h}_{t-1},\boldsymbol{\theta}),
\end{align}
where $a$ is the \textit{Rectified Linear Unit} (ReLU) activation function \cite{relu} applied element-wise; $\textbf{w}$, $\textbf{U}$, $\textbf{b}$, and the parameters $\boldsymbol{\theta}$ of the LSTM are trainable with gradient descent methods; and $\textbf{h}_{t-k}$ is the zero vector. We optimize $f$ using the Root Mean Square Error (RMSE) between the target $x_t$, and predicted value $\hat{x}_t$.
When training $f$ we regularize the model by having a historical window size sequentially increased from $1$ to $k$ as used in language modeling \cite{lstm, graves}.
\subsection{CGM Data Partitions}
As shown in Fig. \ref{fig:flowdiagram}, CGM data from 50 patients and their corresponding periods (14 consecutive CGM days) are partitioned into two separate sets; a training set of 35 randomly selected patients for training of the population-based LSTM and the remaining 15 patients as a test set for patient-specific models.\\
\indent
For each of the 15 patients, their period of 14 CGM days is divided into two 7-day parts. The latter 7 days are used for testing all developed models and the first 7 days are used for training. The training data is then further restricted into either 1, 3, or 7 preceding days such that the test data is contiguous with the preceding training days (dark blue boxes in Fig. \ref{fig:flowdiagram}). These restrictions are made to examine whether increasingly added CGM days will have an impact and improve the performance of the patient-specific models.
\subsection{Data Processing \& Inclusion Criteria}
The CGM data is normalized to have zero mean and a variance of one for all models, i.e. standardizing training data and applying equivalent standardization with training mean and standard deviation on the test data. When training the LSTMs, only historical windows with no missing values are used. At test time, windows with missing values are discarded if they fulfill at least one of two conditions: 1) Input window contains more than 4 missing values within the last hour, and 2) Most recent value is missing. Missing values of remaining windows are linearly interpolated. All missing values are linearly interpolated when fitting ARIMA. Only samples, $x_t$, at which all models have made predictions are considered when evaluating on the test set.
\subsection{Population-based LSTM (Pretraining)}
Taking the population block into account in Fig. \ref{fig:flowdiagram}, the population training set is fed to the population-based LSTM (blue box), also considered as \textit{pretraining} for the patient-specific LSTMs. In this context, a sliding window (using 20 minute steps) of BG values are used as input in batches of 64 windows.\\
\indent
Bayesian optimization (BO) implemented in \textit{SigOpt} \cite{sigopt} is used to efficiently explore a predefined bounded hyperparameter space for LSTMs, dropout level of the MLP layer, and historical window size. 15 distinct LSTM networks (experiments) with the outlined parameters suggested by BO are then trained on 30 patients and evaluated on a validation set of 5 patients. The architecture and configuration corresponding to the best performing population-based LSTM are found in terms of RMSE on the validation set. Networks are optimized by the \textit{Rectified Adam} (RAdam) gradient descent method using default parameters as proposed by \cite{radam}.
\subsection{Patient-specific LSTMs}
The best performing LSTM and its configuration found by BO are used to produce patient-specific LSTMs based on the patient-specific CGM data. To evaluate whether any knowledge from the population-based LSTM can be transferred to patient-specific LSTMs, we evaluate two conditions.\\
\indent
In the first case, we use TL to finetune the population-based LSTM to individual patients (orange box), i.e. learnable parameters of the population-based LSTM are transferred and finetuned with patient-specific training data (1, 3, and 7 CGM days). Given the limited data available on a patient level, a sliding window with 5 minute steps is used while 90\% and 10\% of the data is partitioned into training and validation sets, respectively. Networks are finetuned with Stochastic Gradient Descent (SGD) (learning rate = $10^{-4}$, momentum = 0.9) rather than the adaptive learning rate method RAdam to mitigate catastrophic forgetting.\\
\indent
For the second case, patient-specific LSTMs with the same hyperparameters and patient-specific data as in the first condition are used (green box) but with their network weights randomly initialized rather than transferred from the population-based LSTM. Networks in the second condition are optimized using RAdam with default parameters (as for the population-based LSTM).
\subsection{Scheduling \& Early Stopping}
All LSTMs use a factor-10 reduce-on-plateau learning rate schedule, i.e. learning rate is reduced by a factor of 10 whenever no improvement is obtained within 10 epochs, hence reducing oscillations near local optima. Similarly, in order to prevent overfitting, \textit{early stopping} is applied when no improvement is observed on a held-out validation set within 20 epochs.
\subsection{Patient-specific ARIMA}
For the patient-specific ARIMA models, \verb|AutoARIMA| \cite{pmdarima} is able to fit a suitable model for every case of CGM training day(s) available (1, 3, or 7) for each specific patient (red box). Thus, a total 45 ARIMA models are fitted. The parameter search ranges for $d$, $p$, and $q$ in the \verb|AutoARIMA| are presented in Fig. \ref{fig:flowdiagram} allowing larger ARIMA models to be examined  (i.e. taking up to 2 hours of historical BG values and their errors into account). \verb|AutoARIMA| conducts differencing tests to determine $d$, fit multiple models within a given range for $p$ and $q$, followed by selecting the best model based on the lowest Akaike Information Criterion \cite{pmdarima}. The best performing model parameters are evaluated on the test data. In practice, \verb|AutoARIMA| is implemented by the \verb|pdmarima| Python module and the models are fitted using Newton's method.
\subsection{Prediction Horizon}
Prediction horizons (PHs) of 15, 30, 45, 60, and 90 min into the future are tested, i.e. all models predict 18 BG values into the future (90 min) conditioned on the preceding historical window. \\
\indent
The LSTM models are trained using \textit{teacher forcing} \cite{teacher_forcing}, in which during training the model receives the ground truth output at time $t$ as input at time $t+1$.
\begin{figure}[t]
    \centering
    \includegraphics[scale=0.35]{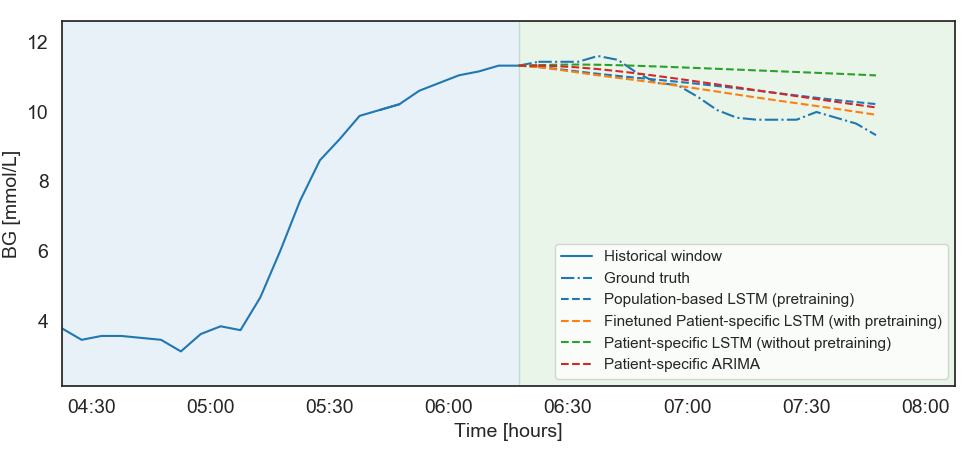}
    \vspace*{-7mm}
    \caption{Example of 90 minutes of predictions for a all the examined models where historical data is followed by ground truth BG values.}
    \label{fig:prediction_illustration}
\end{figure}
For testing, we sequentially predict the next BG value and update the input sequence with the computed prediction in an autoregressive manner. To improve performance of the LSTMs, the \textit{ensemble} BG prediction acquired from 3 different seeds is computed as the final prediction, i.e. the average of 3 separately predicted BG values. \\
\indent
For the ARIMA models, we similarly predict the next BG sequentially and update the input sequence with the computed prediction in an autoregressive manner during testing.
\begin{table*}[t]
\caption{Tables for all the model performances in terms of mean mae and rmse with corresponding standard error of the mean based on the 7 CGM test days of the 15 patients (CGM training days = 7).}
\resizebox{\textwidth}{!}{%
\begin{tabular}{@{}llllll@{}}
\toprule
\multicolumn{1}{c}{\textbf{Model}} & \multicolumn{5}{c}{\textbf{MAE (Standard Error) {[}mmol/L{]}}} \\ \midrule
\textbf{} & \multicolumn{1}{c}{\textbf{15 min}} & \multicolumn{1}{c}{\textbf{30 min}} & \multicolumn{1}{c}{\textbf{45 min}} & \multicolumn{1}{c}{\textbf{60 min}} & \multicolumn{1}{c}{\textbf{90 min}} \\
\textbf{Population-based LSTM (pretraining)} & 0.4392 $\pm$ 0.0282 & 0.8219 $\pm$ 0.0507 & 1.1741 $\pm$ 0.0738 & 1.4919 $\pm$ 0.0980 & 1.9996 $\pm$ 0.1403 \\
\textbf{Finetuned Patient-specific LSTM (with pretraining)} & 0.4406 $\pm$ 0.0297 & 0.8261 $\pm$ 0.0547 & 1.1826 $\pm$ 0.0804 & 1.5049 $\pm$ 0.1066 & 2.0170 $\pm$ 0.1510 \\
\textbf{Patient-specific LSTM (without pretraining)} & 0.4753 $\pm$ 0.0330 & 0.8788 $\pm$ 0.0598 & 1.2406 $\pm$ 0.0842 & 1.5607 $\pm$ 0.1071 & 2.0669 $\pm$ 0.1431 \\
\textbf{Patient-specific ARIMA} & 0.4569 $\pm$ 0.0306 & 0.8587 $\pm$ 0.0558 & 1.2307 $\pm$ 0.0817 & 1.5721 $\pm$ 0.1082 & 2.1241 $\pm$ 0.1542 \\
\textbf{Last Observation Carried Forward} & 0.5650 $\pm$ 0.0356 & 0.9925 $\pm$ 0.0654 & 1.3648 $\pm$ 0.0931 & 1.6901 $\pm$ 0.1185 & 2.2097 $\pm$ 0.1617 \\ \bottomrule
\end{tabular}} 
\label{tab:res}
\resizebox{\textwidth}{!}{%
\begin{tabular}{@{}llllll@{}}
\toprule
\multicolumn{1}{c}{\textbf{Model}} & \multicolumn{5}{c}{\textbf{RMSE (Standard Error) {[}mmol/L{]}}} \\ \midrule
 & \multicolumn{1}{c}{\textbf{15 min}} & \multicolumn{1}{c}{\textbf{30 min}} & \multicolumn{1}{c}{\textbf{45 min}} & \multicolumn{1}{c}{\textbf{60 min}} & \multicolumn{1}{c}{\textbf{90 min}} \\
\textbf{Population-based LSTM (pretraining)} & 0.6492 $\pm$ 0.0449 & 1.1557 $\pm$ 0.0753 & 1.6191 $\pm$ 0.1067 & 2.0317 $\pm$ 0.1397 & 2.6718 $\pm$ 0.1974 \\
\textbf{Finetuned Patient-specific LSTM (with pretraining)} & 0.6493 $\pm$ 0.0459 & 1.1570 $\pm$ 0.0776 & 1.6218 $\pm$ 0.1107 & 2.0364 $\pm$ 0.1446 & 2.6796 $\pm$ 0.2016 \\
\textbf{Patient-specific LSTM (without pretraining)} & 0.6961 $\pm$ 0.0512 & 1.2374 $\pm$ 0.0882 & 1.7101 $\pm$ 0.1195 & 2.1187 $\pm$ 0.1469 & 2.7415 $\pm$ 0.1881 \\
\textbf{Patient-specific ARIMA} & 0.6782 $\pm$ 0.0483 & 1.2158 $\pm$ 0.0821 & 1.7013 $\pm$ 0.1161 & 2.1358 $\pm$ 0.1500 & 2.8284 $\pm$ 0.2098 \\
\textbf{Last Observation Carried Forward} & 0.8023 $\pm$ 0.0528 & 1.3761 $\pm$ 0.0925 & 1.8662 $\pm$ 0.1284 & 2.2876 $\pm$ 0.1610 & 2.9442 $\pm$ 0.2166 \\ \bottomrule
\end{tabular}}
\end{table*}
%
Fig. \ref{fig:prediction_illustration} illustrates 90 minutes of predictions for all the examined models where historical data is followed by ground truth BG values. \\ 
\begin{figure}[h]
    \centering
    \includegraphics[scale=0.43]{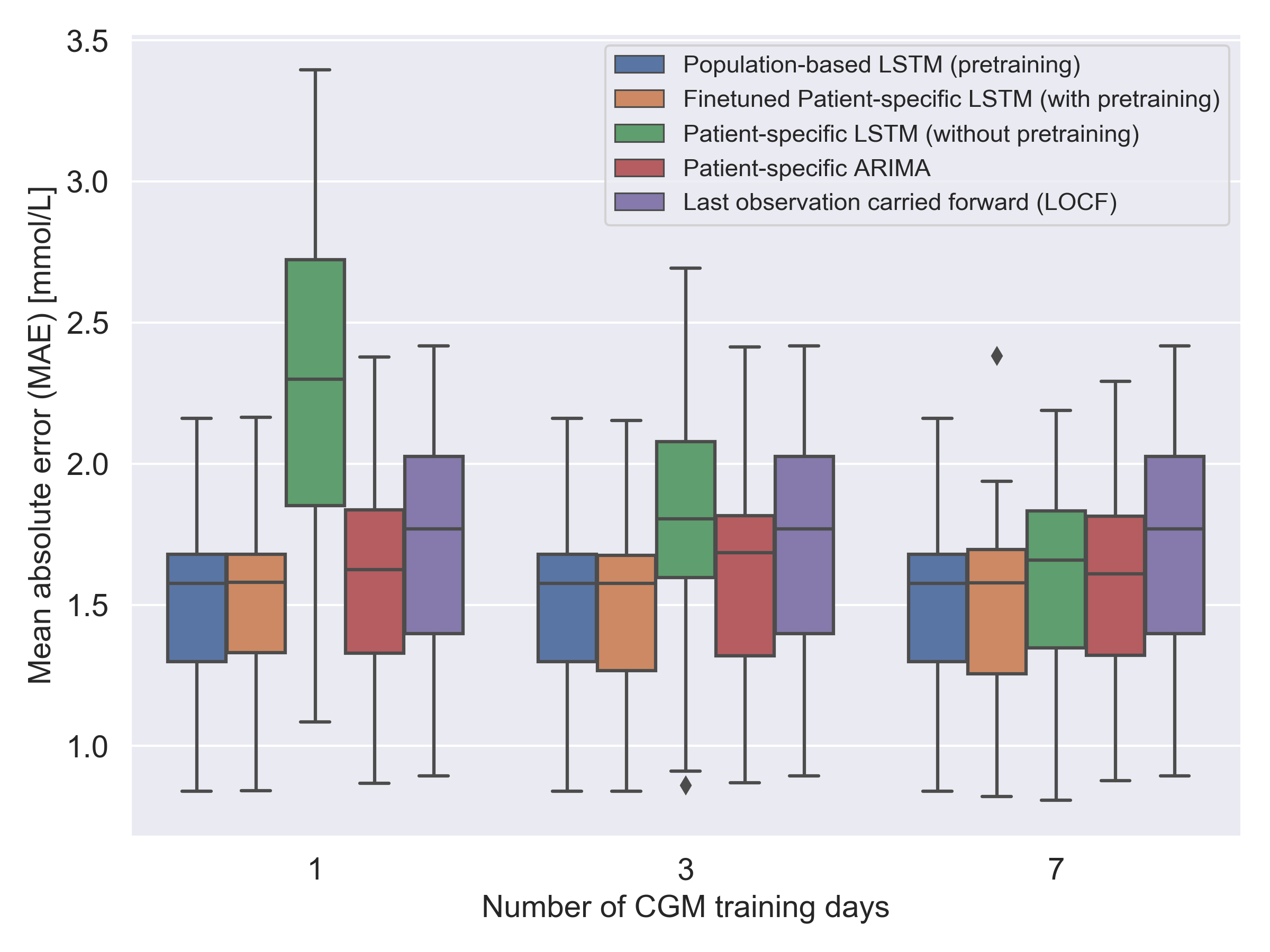}
    \vspace*{-7mm}
    \caption{MAE [mmol/L] for the 15 test patients as function of provided CGM training days (PH = 60 min) for all the models illustrated as boxplots.}
    \label{fig:results1}
\end{figure} 
\begin{figure}[h]
    \centering
    \includegraphics[scale=0.43]{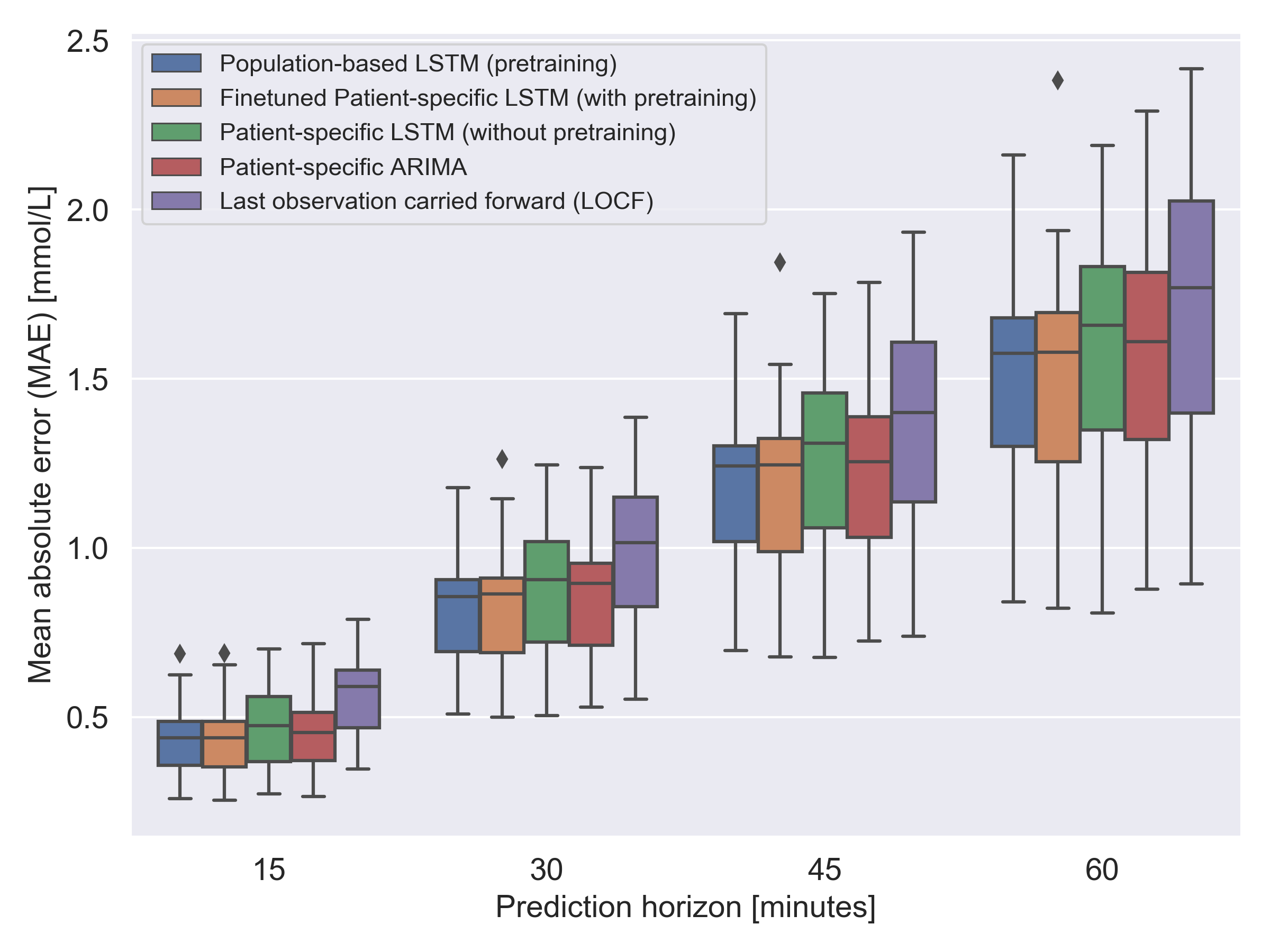}
    \vspace*{-7mm}
    \caption{MAE [mmol/L] for the 15 test patients as function of different PHs (CGM training days = 7) illustrated as boxplots.}
    \label{fig:results2}
\end{figure}
\indent
As performance metrics, the Mean Absolute Error (MAE) (Eq. \ref{eq:mae}) and RMSE (Eq. \ref{eq:rmse2}) are used for model comparison.

\begin{equation}
    \text{MAE}\Big(\textbf{x}_{j}^{(s)}, \hat{\textbf{x}}^{(s)}_{j}\Big) = \frac{1}{n}\sum_{i=1}^{n}\Big|x^{(s)}_{i,j}-\hat{x}^{(s)}_{i,j}\Big|
    \label{eq:mae}
\end{equation}
\begin{align}
    \text{RMSE}\Big(\textbf{x}^{(s)}_{j}, \hat{\textbf{x}}^{(s)}_{j}\Big) = \sqrt{\frac{1}{n}\sum_{i=1}^{n}\Big(x^{(s)}_{i,j}-\hat{x}^{(s)}_{i,j}\Big)^2}
    \label{eq:rmse2}
\end{align}
Here, $n$ represents the number of predictions, $s$ the considered subject and $j$ is the index of PH ranging from 1 to 18.

\section{RESULTS \& DISCUSSION}
The best performing LSTM model found by BO is defined by the following specifications: 2 layers of bidirectional LSTM with a hidden state of size 64, no dropout on the MLP layer (64 hidden units), and 2 hour historical window as input sequence. \verb|AutoARIMA| generally found small models to perform best, with the most commonly occurring complexity being $d=4$, $p=1$, and $q=2$.\\
\indent
Tab. \ref{tab:res} shows all the model performances in terms of mean MAE and RMSE with corresponding standard error of the mean based on the 7 CGM test days of the 15 patients (CGM training days = 7). Additionally, to ground our results from other methods, Last Observation Carried Forward (LOCF) is also included which applies the last available BG observation as its future prediction for all times. \\
\indent
In Tab. \ref{tab:res}, the overall best performing model is the population-based LSTM with MAE ranging between 0.4392 $\pm$ 0.0282 to 1.9996 $\pm$ 0.1403 [mmol/L] for PH = 15 and 90 min, respectively. This is followed by the finetuned patient-specific LSTM both in terms of MAE and RMSE. Patient-specific ARIMA performs slightly better than patient-specific LSTM until PH = 45 min whereas for PH = 60 and 90 min patient-specific LSTM achieves a modestly better performance. As expected, LOCF performs worse than the other models as it does not benefit from the historical trends other than the last BG value. Furthermore, within the investigated models exploiting CGM dynamics, we deal with small performance differences. Fig. \ref{fig:results1} shows the MAE for all models (as boxplots) across the 15 patients as a function of available training CGM days of 1, 3, and 7 having PH fixed to 60 min. \\ 
\indent
Given the described setup in this study, we find that the population-based LSTM (blue) performs most favorable for STBG prediction as the model does not rely on patient-specific CGM training days available. This implies that learning from a population can be beneficial on patient level, thus generalizing well when applied on the unseen patients considered for STBG prediction. From Fig. \ref{fig:results1} it is further clear that performance is patient-dependent given the width of the boxplots. \\
\indent
The finetuned patient-specific LSTM (orange) performs similarly to the population-based LSTM. In other words, the added patient-specific CGM data (7 CGM days) to a pretrained population based LSTM (TL) do not improve predictive performance on individual patients. In this context, there are two possible explanations of the given outcome: 1) More patient-specific CGM data with longer periods are required for training purposes of the LSTMs in order to improve performance on patient level, and 2) TL is known for being difficult to deploy in practice e.g. sub-optimal local minima, slow convergence, and catastrophic forgetting indicating undesired pitfalls during training. \\
\indent
On the other hand, considering the patient-specific LSTM (green), the number of available CGM days for each specific patient have an impact. The model transcends LOCF with 7 training days, initially being the worst performing model with only one training day (see Fig. \ref{fig:results1}). This suggests that more patient-specific historical data will benefit the performance and enable learning individual trends. \\
\indent
The patient-specific ARIMA models (red) perform well in general regardless of available CGM training days. Nevertheless, it has a larger midspread than the population-based LSTM (blue). Hence, the linear models perform comparably with the more complex LSTM-based models. Simultaneously, if only limited CGM data is available (in this case up to 7 CGM days), a patient-specific ARIMA would be a promising alternative to the patient-specific LSTM. \\
\indent
Fig. \ref{fig:results2} visualizes the change of MAE as a function of different PHs, with the number of CGM training days fixed to 7. As expected, MAE is increased by extended PH for all the models. Simultaneously, it also indicates for the LOCF approach that a short PH of e.g. 15 minutes might still provide some insight of the BG level depending on the task. \\
\indent
Limitations of this study includes the relatively small number of patients. Considering a total sample of 50 patients may show indications, but more patients and CGM data are required to investigate consistency and robustness to the developed models. \\
\indent
Another limitation is unspecified patient information, i.e. we do not know the diabetes type, demographics or treatment regimens, all of which are known to impact BG. In particular, T1D and T2D BG dynamics are very different and their dissimilar effects might distort the underlying signal in the CGM data. \\
\indent
Future studies should focus on a specific cohort of patients with known treatment regimens, demographics and if possible other data sources (carbohydrates, exercise, dose size). In particular, examining population- and patient-tailored STBG prediction models aimed towards specific group of patients. Moreover, the problem could also be approached as a classification problem as opposed to regression. In this context, targeted towards hypoglycemic event detection including e.g. expert-dependent features acquired from the CGM data, a strategy successfully used in other tasks \cite{mohebbi, mohebbi2, adherence}. Ideally, future solutions could be based on different hybrid models which are dynamically deployed depending on data availability and quality. 
\section{CONCLUSION}
In this study, we explore the effectiveness and potential of using LSTM-based RNNs for STBG prediction in contrast to a linear model such as ARIMA. We find that a population-based LSTM model generalizes well to individual patients, being the best performing model in all PH cases. Adding up to 7 CGM training days in the finetuned patient-specific LSTM (TL) do not seem to suffice/improve performance. This indicates the need for further investigation and more patient-specific CGM data with longer periods. The patient-specific LSTM does not perform well when only one day of CGM data are available. On the other hand, it benefits by added patient-specific training days performing similar to the patient-specific ARIMA with 7 days of CGM data available. \\
\indent
Given the outcome of this study, advanced predictive models could be deployed when designing decision support tools with BG forecasting capabilities. However, in some situations the conventional linear ARIMA approach performs as well
or better, in particular when little CGM data are available. A future solution may be a hybrid predictive approach dynamically deploying different models depending on data availability and quality.


\bibliographystyle{IEEEtran}
\bibliography{refs}

\begin{thebibliography}{10}
\providecommand{\url}[1]{#1}
\csname url@samestyle\endcsname
\providecommand{\newblock}{\relax}
\providecommand{\bibinfo}[2]{#2}
\providecommand{\BIBentrySTDinterwordspacing}{\spaceskip=0pt\relax}
\providecommand{\BIBentryALTinterwordstretchfactor}{4}
\providecommand{\BIBentryALTinterwordspacing}{\spaceskip=\fontdimen2\font plus
\BIBentryALTinterwordstretchfactor\fontdimen3\font minus
  \fontdimen4\font\relax}
\providecommand{\BIBforeignlanguage}[2]{{%
\expandafter\ifx\csname l@#1\endcsname\relax
\typeout{** WARNING: IEEEtran.bst: No hyphenation pattern has been}%
\typeout{** loaded for the language `#1'. Using the pattern for}%
\typeout{** the default language instead.}%
\else
\language=\csname l@#1\endcsname
\fi
#2}}
\providecommand{\BIBdecl}{\relax}
\BIBdecl

\bibitem{tech}
S.~Grock, J.-h. Ku, J.~Kim, and T.~Moin, ``A review of technology-assisted
  interventions for diabetes prevention,'' \emph{Current diabetes reports},
  vol.~17, no.~11, p. 107, 2017.

\bibitem{cobelli}
C.~Zecchin, A.~Facchinetti, G.~Sparacino, and C.~Cobelli, ``How much is
  short-term glucose prediction in type 1 diabetes improved by adding insulin
  delivery and meal content information to cgm data? a proof-of-concept
  study,'' \emph{Journal of diabetes science and technology}, vol.~10, no.~5,
  pp. 1149--1160, 2016.

\bibitem{consensus}
T.~Battelino, T.~Danne, R.~M. Bergenstal, S.~A. Amiel, R.~Beck, T.~Biester,
  E.~Bosi, B.~A. Buckingham, W.~T. Cefalu, K.~L. Close \emph{et~al.},
  ``Clinical targets for continuous glucose monitoring data interpretation:
  Recommendations from the international consensus on time in range,''
  \emph{Diabetes care}, p. dci190028, 2019.

\bibitem{mohebbi}
A.~Mohebbi, J.~M. Tarp, M.~L. Jensen, S.~Puthusserypady, E.~Hachmann-Nielsen,
  H.~Bengtsson, and M.~M{\o}rup, ``Fast assessment of glycemic control based on
  continuous glucose monitoring data,'' in \emph{2019 41st Annual International
  Conference of the IEEE Engineering in Medicine and Biology Society
  (EMBC)}.\hskip 1em plus 0.5em minus 0.4em\relax IEEE, 2019, pp. 7185--7188.

\bibitem{mohebbi2}
A.~Mohebbi, T.~B. Arad{\'o}ttir, A.~R. Johansen, H.~Bengtsson, M.~Fraccaro, and
  M.~M{\o}rup, ``A deep learning approach to adherence detection for type 2
  diabetics,'' in \emph{2017 39th Annual International Conference of the IEEE
  Engineering in Medicine and Biology Society (EMBC)}.\hskip 1em plus 0.5em
  minus 0.4em\relax IEEE, 2017, pp. 2896--2899.

\bibitem{ARIMA}
G.~E. Box, G.~M. Jenkins, G.~C. Reinsel, and G.~M. Ljung, \emph{Time series
  analysis: forecasting and control}.\hskip 1em plus 0.5em minus 0.4em\relax
  John Wiley \& Sons, 2015.

\bibitem{stbg_ref}
S.~Fiorini, C.~Martini, D.~Malpassi, R.~Cordera, D.~Maggi, A.~Verri, and
  A.~Barla, ``Data-driven strategies for robust forecast of continuous glucose
  monitoring time-series,'' in \emph{2017 39th Annual International Conference
  of the IEEE Engineering in Medicine and Biology Society (EMBC)}.\hskip 1em
  plus 0.5em minus 0.4em\relax IEEE, 2017, pp. 1680--1683.

\bibitem{AI}
Y.~LeCun, Y.~Bengio, and G.~Hinton, ``Deep learning,'' \emph{nature}, vol. 521,
  no. 7553, p. 436, 2015.

\bibitem{lstm}
S.~Hochreiter and J.~Schmidhuber, ``Long short-term memory,'' \emph{Neural
  computation}, vol.~9, no.~8, pp. 1735--1780, 1997.

\bibitem{graves}
A.~Graves, ``Generating sequences with recurrent neural networks,'' \emph{arXiv
  preprint arXiv:1308.0850}, 2013.

\bibitem{rcnn}
K.~Li, J.~Daniels, C.~Liu, P.~Herrero-Vinas, and P.~Georgiou, ``Convolutional
  recurrent neural networks for glucose prediction,'' \emph{IEEE Journal of
  Biomedical and Health Informatics}, 2019.

\bibitem{transfer}
A.~Zhang, H.~Wang, S.~Li, Y.~Cui, Z.~Liu, G.~Yang, and J.~Hu, ``Transfer
  learning with deep recurrent neural networks for remaining useful life
  estimation,'' \emph{Applied Sciences}, vol.~8, no.~12, p. 2416, 2018.

\bibitem{relu}
X.~Glorot, A.~Bordes, and Y.~Bengio, ``Deep sparse rectifier neural networks,''
  in \emph{Proceedings of the fourteenth international conference on artificial
  intelligence and statistics}, 2011, pp. 315--323.

\bibitem{sigopt}
I.~Dewancker, M.~McCourt, and S.~Clark, ``Bayesian optimization primer,'' 2015.

\bibitem{radam}
L.~Liu, H.~Jiang, P.~He, W.~Chen, X.~Liu, J.~Gao, and J.~Han, ``On the variance
  of the adaptive learning rate and beyond,'' \emph{arXiv preprint
  arXiv:1908.03265}, 2019.

\bibitem{pmdarima}
\BIBentryALTinterwordspacing
T.~G. Smith \emph{et~al.}, ``{pmdarima}: Arima estimators for {Python},''
  2017--, [Online; accessed 10.11.2019]. [Online]. Available:
  \url{http://www.alkaline-ml.com/pmdarima}
\BIBentrySTDinterwordspacing

\bibitem{teacher_forcing}
R.~J. Williams and D.~Zipser, ``A learning algorithm for continually running
  fully recurrent neural networks,'' \emph{Neural computation}, vol.~1, no.~2,
  pp. 270--280, 1989.

\bibitem{adherence}
D.~Thyde, A.~Mohebbi, H.~Bengtsson, M.~L.~Jensen, and M.~Mørup, ``Machine
  learning based adherence detection of type 2 diabetes patients on once-daily
  basal insulin injections,'' \emph{Journal of diabetes science and
  technology}, 2019.

\end{thebibliography}

\end{document}